\documentclass[apj]{emulateapj}
\usepackage{color}
\usepackage{epsf}
\usepackage{ulem}
\usepackage{epsfig}
\usepackage{graphicx}

\def\h0 {$h_0$=71 km s$^{-1}$ Mpc$^{-1}$}
\def\ergs { erg s$^{-1}$}
\def\ga {$\Gamma_{\rm 2-10keV}$}
\def\me {$\dot m_{\rm E}$}
\def\lb {$L_{\rm bol}$}

\newcommand{\be}{\begin{equation}}
\newcommand{\ee}{\end{equation}}
\newcommand{\ce}{\ifmmode {\cal E} \else ${\cal E}$\ \fi}
\newcommand{\kms}{\ifmmode {\rm km\ s}^{-1} \else km s$^{-1}$\ \fi}
\newcommand{\tes}{\ifmmode \tau_{\rm es} \else $\tau_{\rm es}$\ \fi}
\newcommand{\tk}{\ifmmode \tau_{\rm K} \else $\tau_{\rm K}$\ \fi}
\newcommand{\vfwhm}{\ifmmode V_{\mbox{\tiny FWHM}} \else
            $V_{\mbox{\tiny FWHM}}$\fi}
\newcommand{\msun}{\ifmmode M_{\odot} \else $M_{\odot}$\ \fi}
\newcommand{\afe}{\ifmmode {\mathcal A_{\rm Fe}} \else${\mathcal A_{\rm Fe}}$\ \fi}

\newcommand{\ledd}{\ifmmode L_{\rm Edd} \else $L_{\rm Edd}$\ \fi}
\newcommand{\lx}{\ifmmode L_{\rm 2-10keV} \else  $L_{\rm 2-10keV}$\ \fi}
\newcommand{\hb}{\ifmmode H\beta \else H$\beta$\ \fi}
\newcommand{\mbh}{\ifmmode M_{\rm BH}  \else $M_{\rm BH}$\ \fi}
\newcommand{\lv}{\ifmmode \lambda L_{\lambda}(5100\AA) \else $\lambda L_{\lambda}(5100\AA)$\ \fi}

\newcommand{\civ}{C {\sc iv}\ }

\newcommand{\mgii}{Mg {\sc ii}\ }

\def\ariel5{{\it Ariel 5}\ }

\def\xmm{{\it XMM-Newton}\ }
\def\chandra{{\it Chandra}\ }

\def\heao1{{\it HEAO~1}\ }

\shorttitle{AGN cosmology} \shortauthors{Zhou
\& Zhao}

\begin{document}
\title{Hard X-ray photon index as an indicator of
bolometric correction in active galactic nuclei}

\author{Xin-Lin Zhou\altaffilmark{1,2}, Yong-Heng Zhao\altaffilmark{1,2}}
\altaffiltext{1}{Key Laboratory of Optical Astronomy, National
Astronomical Observatories, Chinese Academy of Sciences, Beijing,
100012, China} \altaffiltext{2} {National Astronomical
Observatories, Chinese Academy of Sciences, Beijing, 100012, China}

\email{zhouxl@nao.cas.cn}


\begin{abstract}
We propose the rest-frame 2-10 keV photon index, \ga, acting as an
indicator of the bolometric correction, \lb/$L_{\rm 2-10keV}$ (where
\lb~ is the bolometric luminosity and $L_{\rm 2-10keV}$ is the
rest-frame 2-10 keV luminosity), in radio-quiet active galactic
nuclei (AGNs). Correlations between \ga~ and both
bolometric correction and Eddington ratio are presented, based on
simultaneous X-ray, UV, and optical observations of reverberation
-mapped AGNs. These correlations can be compared with those for
high-redshift AGNs to check for any evolutionary effect. Assuming no
evolutionary effect in AGNs' spectral properties, together with the
independent estimates of $L_{\rm 2-10keV}$, the bolometric
correction, Eddington ratio, and black hole (BH) mass can all be estimated
from these correlations for high-redshift AGNs, with the mean
uncertainty of a factor of 2-3. If there are independent estimates
of BH masses, \ga~ for high-redshift AGNs can be used to
determine their true \lb~ and $L_{\rm 2-10keV}$, and in conjunction
with the redshift, can be potentially used to place constraints on
cosmology by comparison with the rest-frame 2-10 keV flux. We find
that the true $L_{\rm 2-10keV}$ estimated from \ga~ for the
brightest Type I AGNs with $z<1$ in the Lockman Hole is generally in
agreement with the observed $L_{\rm 2-10keV}$.
 However, there are still many uncertainties, such as the
accurate determination of the intrinsic \ga~ for distant AGNs and
the large uncertainty in the luminosities obtained, which call for
significant further study before ``AGN cosmology'' can be considered a
viable technique.

\end{abstract}


\section{Introduction}
Active galactic nuclei (AGNs) and quasars, as the most powerful
long-lasting celestial bodies, are believed to be powered by the
accretion of gas onto supermassive black holes (BHs). The
cosmological  applications of AGNs have long been explored since
they have been detected reaching the redshift of $\sim6.4$ (Fan et
al. 2006). Gunn \& Peterson (1965) proposed, using the Ly$\alpha$
resonance absorption in the spectra of distant quasars as a probe
for the neutral hydrogen density in the intergalactic medium at high
redshift  (see the review by Fan et al. 2006). Baldwin (1977)
proposed the equivalent width of the \civ line as a luminosity
indicator. Since then, many authors studied the correlations between
the equivalent width of emission lines and the continuum
luminosities (e.g., Shang et al. 2003; Baskin \& Laor 2004; Zhou \&
Wang 2005). However, these correlations show large scatters
with uncertain slopes. It is difficult for the equivalent width of
emission lines to work as a luminosity indicator.

Although it is difficult to find a simple luminosity indicator for
AGNs, one can estimate the true luminosity of an AGN in a distant
independent way. This is intriguing since AGNs may potentially
become cosmological probes.
 The bolometric luminosity, $L_{\rm bol}$, of an AGN
depends on its BH mass, $M_{\rm BH}$, and Eddington ratio,
 $\dot m_{\rm E}\equiv L_{\rm
bol}/L_{\rm Edd}$, where $L_{\rm Edd}$ is the Eddington luminosity
for an AGN's $M_{\rm BH}$.  It has been found that the rest-frame
2-10 keV photon index, $\Gamma_{\rm 2-10 keV}$, is an indicator of
\me ~(Lu \& Yu 1999; Wang et al. 2004; Kelly et al. 2007; Zhou et
al. 2007; Shemmer et al. 2008, hereafter S08). If $M_{\rm BH}$ can be
estimated from a method which is independent of the luminosity, e.g.,
from the X-ray variability amplitude (Zhou et al. 2010) or the
stellar velocity dispersion (Tremaine et al. 2002), the bolometric
luminosity of an AGN can be calculated from \ga~ and $M_{\rm BH}$,
under the assumption of no evolutionary effect in AGNs' spectral
properties. If the rest-frame 2-10 keV luminosity, $L_{\rm 2-10
keV}$, can be estimated from \lb, we can derive an AGN's luminosity
distance by comparison with the absorption-corrected 2-10 keV flux.
Therefore, the bolometric correction, $\kappa_{\rm 2-10 keV}$,
defined by the ratio of \lb~ to $L_{\rm 2-10 keV}$, should also be
studied. The previous work on the spectral energy distributions
(SEDs) of AGNs suggested that the spread of $\kappa_{\rm 2-10 keV}$
of individual AGNs is large (Elvis et al. 1994). It was found that
the optical-to-X-ray SED shape depends on the UV luminosity (e.g.,
Yuan et al. 1998) and then the dependence of $\kappa_{\rm 2-10 keV}$
on the luminosity is proposed (Marconi et al. 2004). However,
Vasudevan \& Fabian (2007, hereafter VF07) found that $\kappa_{\rm 2-10 keV}$
is a function of \me, with no clear dependence on luminosity.

Here we propose \ga~ to be an indicator of $\kappa_{\rm 2-10 keV}$.
We calibrate the correlations between \ga/$\kappa_{\rm 2-10 keV}$,
and \ga/$\dot m_{\rm E}$ based on simultaneous X-ray, UV, and optical
observations of low-redshift AGNs with reverberation-based $M_{\rm
BH}$ (Peterson et al. 2004). These correlations can be used to
obtain $M_{\rm BH}$ (together with the independent luminosity), or
the estimates of an AGN's true luminosity (together with the
independent $M_{\rm BH}$). Throughout this Letter, we assume a
cosmology of \h0 , $\Omega_{\rm m}=0.27$, and
$\Omega_{\Lambda}=0.73$.

\section{Sample and Data}

Simultaneous X-ray, UV and optical observations of 29 low-redshift
($z<0.33$) AGNs in Peterson et al.'s (2004) sample obtained from the
\xmm~ EPIC and OM instruments have been reduced by Vasudevan \&
Fabian (2009, hereafter VF09). The purpose of this work is studying \ga~ as an
indicator of $\kappa_{\rm 2-10 keV}$ and \me, but the beaming
emission of radio-loud AGNs may affect the measurements of the
intrinsic \ga. Therefore, we remove radio-loud AGNs 3C 120, 3C 390.3,
and 3C 273 from this sample. We also remove the quasar PG 1411+442
since this object is heavily obscured. Note that there are a few
other sources in the reverberation-mapped sample which, although
they may not be heavily absorbed, may have some kind of complex
absorption. In those cases it is difficult to really know what the
intrinsic \ga~ is for those objects. VF09 used a broken power-law
model to model the 1-8 keV data. This simple approach cannot model
the shape of the hard X-ray continuum precisely. Since the whole
premise of this work rests on getting good, accurate \ga~ values for
this sample, we take \ga~ from the more detailed fits available in
the literature for the X-ray observations. We present results from
27 observations of 25 AGNs in Table 1.

Most of objects in Table 1 are included in VF07's sample. VF09
compared the shape of the simultaneous SEDs of AGNs in Table 1 with
the non-simultaneous SEDs in VF07. The values of \me~ and
$\kappa_{\rm 2-10 keV}$ obtained in VF09 show a similar range to
those in VF07. However, \me~ values in Table 1 on average are a
factor of $\sim0.8$ of VF07's values due to the generally larger
$M_{\rm BH}$ in Peterson et al. (2004), and $\kappa_{\rm 2-10 keV}$
values in Table 1 are on average $\sim1.38$ times larger than VF07's
values.

\section{Results}

\subsection{Photon index and Eddington ratio}

We assume that there is a linear
correlation, $y=\alpha +\beta x$,
with the measurement errors of $\epsilon_{x_i}$ for $x_i$ and
$\epsilon_{y_i}$ for $y_i$. It was shown that the Nukers' estimate
 is an unbiased slope estimator for the linear regression (Tremaine et al. 2002).
The Nukers' estimate is based on minimizing:

\be \chi^2\equiv \sum_{i=1}^N{(y_i-\alpha - \beta x_i)^2\over
\epsilon_{yi}^2+\beta^2\epsilon_{xi}^2}.
 \label{eq:chisq} \ee

Figure 1(a) shows the correlation between \ga~ and \me~ for the
present sample. The correlation is significant, with a Spearman's
coefficient of 0.55. The Spearman's probability associated with this
coefficient is about 0.3\%. We then apply the Nukers' estimate using
the {\it fitexy} routine (Press et al. 1992)
 to derive the correlation between \ga~ and \me,
 \be \label{eq:GaEdd}\Gamma_{\rm 2-10 keV} = (0.31\pm0.02){\rm
log}\dot m_{\rm E} +(2.11\pm0.02). \ee

The minimum $\chi^2$ per degree of freedom is 21.8, indicating that
either the uncertainties on $\Gamma_{\rm 2-10 keV}$ are
underestimated or there is an intrinsic dispersion in the
correlation.  To account for the intrinsic dispersion
 in the $M_{\rm BH}-\sigma^2_{\rm rms}$ relation, we replace $\epsilon_{yi}$ by
$(\epsilon_{yi}^2+\epsilon_0^2)^{1/2}$, where $\epsilon_0$
represents the intrinsic dispersion; $\epsilon_0$ is adjusted so
that the value of $\chi^2$ per degree of freedom is unity. This
procedure is preferable if the individual error estimates,
$\epsilon_{y_i}$, are reliable (Tremaine et al. 2002).
 Adding an intrinsic dispersion of 0.51 dex decreases the value
  of $\chi^2$ per degree of freedom to unity and gives the best-fit
  result:

\be {\rm log} \dot m_{\rm E}=(2.09\pm0.58)\Gamma_{\rm 2-10 keV} -
(4.98\pm1.04), \ee

as shown in Figure 1(a).

\subsection{Photon Index and Bolometric Correction}
Figure 1(b) shows the correlation between \ga~ and $\kappa_{\rm 2-10
keV}$. A Spearman's test returns the correlation coefficient of
0.59, corresponding to the Spearman's probability of 0.1\%. We then
apply the Nukers' estimate to derive the correlation:

\be \label{eq:GaBC} {\rm log} \kappa_{\rm 2-10 keV} =
(2.52\pm0.08)\Gamma_{\rm 2-10 keV} -(3.12\pm0.15). \ee

The minimum $\chi^2$ per degree of freedom is 35, indicating that
either the uncertainties in $\kappa_{\rm 2-10 keV}$ are
underestimated, or there is an intrinsic dispersion in the
correlation. We add an intrinsic dispersion of 0.32 dex to decrease
the value of $\chi^2$ per degree of freedom to unity, which gives
the best-fit result: \be \label{eq:GaBCb} {\rm log} \kappa_{\rm 2-10
keV} = (1.12\pm0.30)\Gamma_{\rm 2-10 keV} -(0.63\pm0.53), \ee

as shown in Figure 1(b).

\section{Discussion}
The reverberation-mapped AGNs used here are a subset of AGNs, which
are skewed toward moderate luminosity Seyfert 1 galaxies.  The
bolometric luminosity ranges from 10$^{42.4}$ \ergs~ to 10$^{46.5}$
\ergs. The Eddington ratio ranges from 0.0004 to 1.13, with a mean
value of $\sim0.16$, and only a few AGNs have low values. This is
roughly in agreement with a large sample of Seyfert 1 galaxies
selected from the Sloan Digital Sky Survey (Figure 12 in Shen et al.
2008).

Equation (2) is in excellent agreement with Equation (1) in S08,
which includes high-luminosity AGNs reaching $z\sim$ 3.2. This
suggests that there is no evidence for the evolution of AGNs' spectral
properties over cosmological time. Many previous studies suggested
that X-ray spectra of high-redshift AGNs (reaching $z\sim6$) remain
similar properties to those of low-redshift ones (e.g., Shemmer et
al. 2005). Nevertheless, the spectral bandpass is affected by the
redshift. The 2-10 keV band in the observed frame will correspond to
energies higher by a factor of (1+$z$) in the rest frame. Thus, the
rest-frame 2-10 keV band should be used for high-redshift AGNs.

The first idea proposed here is that \ga~ can be used to estimate
\me~ and $\kappa_{\rm 2-10 keV}$, which can then be used along with
$L_{\rm 2-10 keV}$ to get $M_{\rm BH}$. A similar possibility was
discussed in Zhou et al. (2007) and S08. S08 used the
luminosity-dependent $\kappa_{\rm 2-10 keV}$ in Marconi et al.
(2004). Note that Marconi et al. (2004) used the nonlinear
$\alpha_{\rm ox}-L_{2500}$ correlation (e.g., Section 3.3 in S08) to
work out their $\kappa_{\rm 2-10 keV}$. Here, we use values of \lb~
determined from the SEDs given by the simultaneous X-ray, UV, and
optical observations. We find a robust correlation between \ga~ and
$\kappa_{\rm 2-10 keV}$. This allows the direct estimate of
$\kappa_{\rm 2-10 keV}$ from \ga. This significantly extends the
work of S08. S08 also argued that this approach can give an estimate
of both \me~ and $M_{\rm BH}$ with mean uncertainties within a
factor of $<3$. This is in good agreement with the intrinsic
dispersion of correlations derived here. Our best-fitting result of
the \ga$-$\me~ correlation, Equation (3), is different from Equation
(2) in S08. We tend to use our result since the correlation based on
reverberation mapped AGNs is more robust.

The second idea proposed here is that \ga~ can be used, along with
an independent measure of $M_{\rm BH}$, to give \lb~ and $L_{\rm
2-10 keV}$. Note that $L_{\rm bol}= \dot m_{\rm E}\times L_{\rm
Edd}$, where $L_{\rm Edd}=1.26\times10^{38}M_{\rm BH}$ erg s$^{-1}$.
Provided the intrinsic \ga~ is known accurately enough from good
quality X-ray data, one can readily get $L_{\rm bol}$ from \ga~ and
$M_{\rm BH}$. One can also estimate $\kappa_{\rm 2-10 keV}$ from
\ga~ based on our Equation (5).  This allows the direct calculation
of $L_{\rm 2-10 keV}$ by using the Equation of $L_{\rm 2-10
keV}=L_{\rm bol}/\kappa_{\rm 2-10 keV}$. Provided the
absorption-corrected rest-frame 2-10 keV flux can be obtained
accurately, the luminosity distance can be determined and the AGN
becomes a potential cosmological probe.

Figure 2 shows the true $L_{\rm 2-10 keV}$ (open circle) estimated
from \ga~ and $M_{\rm BH}$, compared with the observed $L_{\rm 2-10
keV}$ (filled circle) calculated from the X-ray flux for the
brightest Type I AGNs with $z<1$ taken from \xmm~ observations of
the Lockman Hole (Mateos et al. 2005). The true $L_{\rm 2-10 keV}$
is generally in agreement with the observed $L_{\rm 2-10 keV}$, with
a notable outlier of RDS 426A. This object is very heavily absorbed,
with a column density of hydrogen of $10^{22.09}$ cm$^{-2}$ (Mateos
et al. 2005). $M_{\rm BH}$ is derived from the empirical virial
correlation using the \hb or \mgii line and the optical luminosity
available in Lehmann et al. (2001). The $M_{\rm BH}$ estimates
mainly depend on the emission-line width,  and weakly on the optical
continuum luminosity.
 The second idea requires that $M_{\rm BH}$ should be measured from
the method independent of the luminosity. The stellar velocity
dispersion, $\sigma_\ast$, is an $M_{\rm BH}$ estimator independent
of luminosity.
 However, it is difficult to measure $\sigma_\ast$ for
high-redshift AGNs. Furthermore, there was some evidence for the
evolution of the $M_{\rm BH}-\sigma_\ast$ correlation reaching
$z\sim2$ (Trakhtenbrot \& Netzer 2010, and the references therein).
Recently, Zhou et al. (2010) suggested that the X-ray variability
amplitude (so-called excess variance; Nandra et al. 1997) can act as
an indicator of $M_{\rm BH}$. It is also difficult to obtain
high-quality X-ray light curves to validate this technique at high
redshift.

If there are independent estimates of $M_{\rm BH}$ in the future,
correlations presented here can also be used to estimate the true
luminosity of high-redshift AGNs, which will enable them to be used
in a similar way that Type Ia supernovae have been previously used
for cosmology applications. AGNs are so ubiquitous in deep X-ray
surveys, that their numbers provide significant potential for such
work. However, since this method rests on the accurate determination
of the intrinsic \ga~ to get $\dot m_{\rm E}$ and $\kappa_{\rm 2-10
keV}$, it will be difficult to get the intrinsic \ga~ for high-redshift
 AGNs. For example, in the \chandra Deep Fields, only simple
hardness ratios can be calculated for many of AGNs, and the steep
ratios seen for many of these AGNs indicate that many of them could
be heavily absorbed, making it difficult to obtain the intrinsic
\ga~ (Alexander et al. 2003). It is expected that the future
International X-ray Observatory, or any other future instruments,
will really offer the capabilities to determine \ga~ sufficiently
accurately at higher redshift.

Provided the accurate \ga~ for high-redshift AGNs can be obtained,
the issue still remains that an AGN's luminosity varies
significantly over even short timescales. The correlations presented
here can be used for the luminosity estimates, but the uncertainty
inherent in these correlations would be just as big as a stumbling
block to doing cosmology. Assuming that the error of $M_{\rm BH}$ is 0.3
dex, assuming \ga~ comes with an error of $\pm 0.2$, the error of
luminosity estimates from these correlations is $\sim0.37$. This
 is just the lower limit of actual one due to the X-ray
variability, since the X-ray flux-\ga~ correlation exhibits an
inverted behaviour for an AGN (Figure 9 in Ponti et al. 2006). The
error of $M_{\rm BH}$ may be underestimated for high-redshift
objects. This further enlarges the uncertainty in the luminosity
obtained. Therefore, it is difficult to give useful constraints on
cosmology with the AGN data. The refined versions of the
correlations would need to be produced in future with cleaner
objects.

\section{Conclusions}
Correlations between hard X-ray photon index and both bolometric
correction and Eddington ratio are presented, based on simultaneous
X-ray, UV, and optical observations of radio-quiet AGNs with
reverberation-based BH masses. We thus propose X-ray photon index
acting as an indicator of bolometric correction. Assuming no
evolutionary effect of AGNs' spectral properties, together with the
X-ray luminosity, the bolometric correction, Eddington ratio, and BH
mass can all be estimated from these correlations for the
high-redshift AGNs, with the mean uncertainty within a factor of
2-3.

With the BH mass known to be independent of the luminosity, the true
luminosity of an AGN can be calculated from its X-ray photon index
based on the correlations presented here. We find that the true
X-ray luminosity estimated from the photon index for the brightest Type
I AGNs with $z<1$ in the Lockman Hole is generally in agreement with
the observed X-ray luminosity. However, the errors of AGNs'
luminosity estimates depend on the various uncertainties in the
X-ray spectra, variability, and some unknown systematic effects in
the distant BH mass estimates, which is too much to get an accurate
estimate of the cosmology.

\acknowledgments We acknowledge the two referees for many useful
comments to improve the manuscript significantly. James Wicker is
thanked for improving English. This work is supported by the
Guoshoujing Telescope (formerly named the Large Sky Area
Multi-Object Fiber Spectroscopic Telescope - (LAMOST)), which is
funded by the National Development and Reform Commission, operated
and managed by the Key Laboratory of Optical Astronomy, NAOC. This
research has made use of observations obtained with {\it
XMM-Newton}, an ESA science mission with instruments and
contributions directly funded by ESA member states and NASA, USA.



\clearpage

\begin{figure}
\includegraphics[width=11 cm, angle=270]{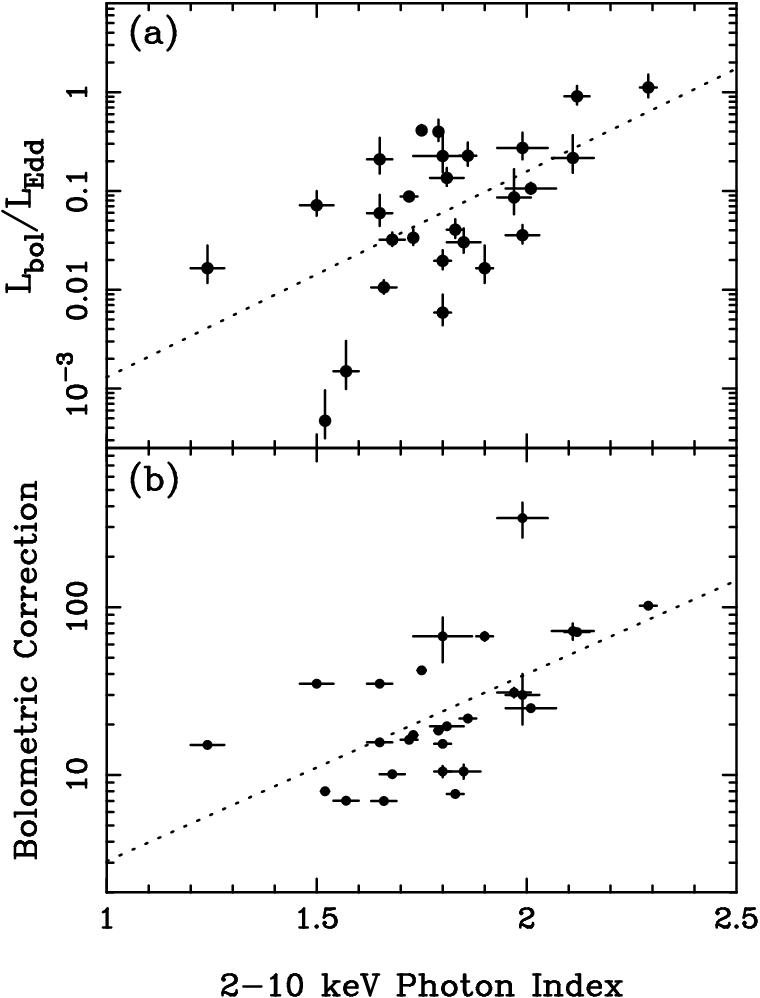}\label{fig1}
\caption  {Panel (a): Correlation between \ga~ and \me. The dotted line denotes the best
fitting result of Equation (3); Panel (b):
Correlation between \ga~ and $\kappa_{\rm 2-10 keV}$. The dotted line denotes the best fitting result of Equation (5).
}
\end{figure}

\begin{figure}
\includegraphics[width=11 cm, angle=270]{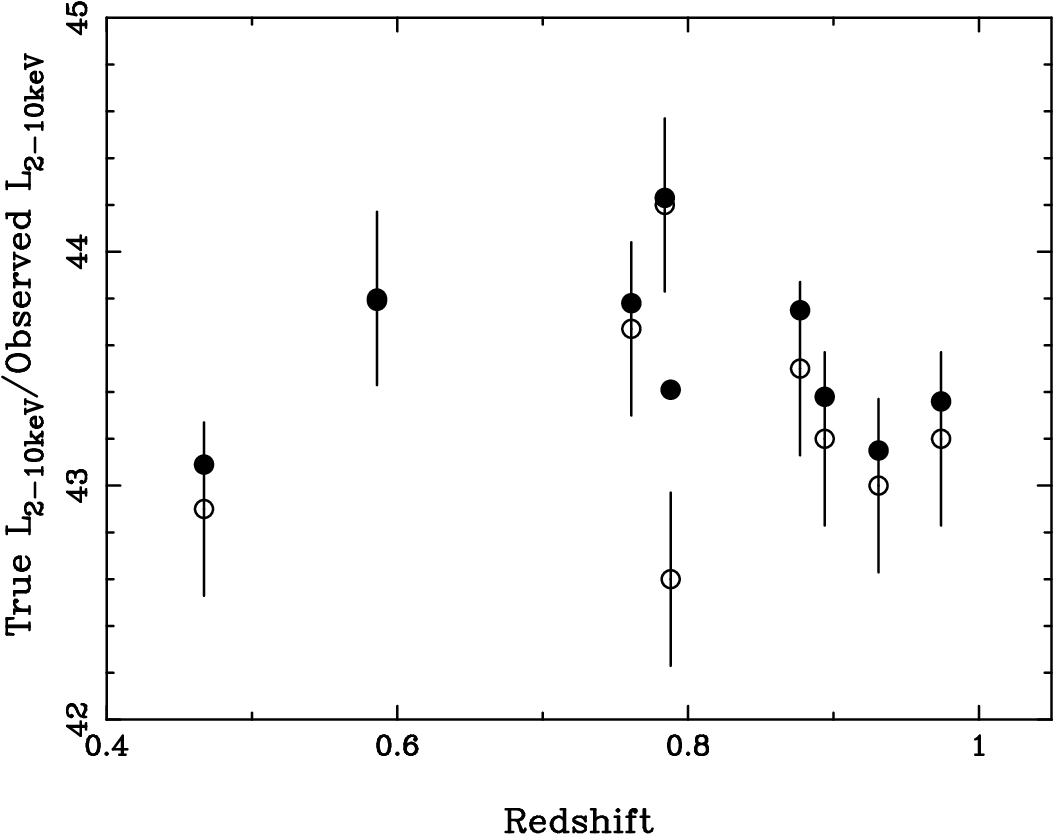}\label{fig2}
\caption  { True $L_{\rm 2-10 keV}$ (open circle) estimated from
\ga~ and $M_{\rm BH}$, compared with the observed $L_{\rm 2-10 keV}$
(filled circle) calculated from the X-ray flux for the brightest
Type I AGNs with $z<1$ taken from \xmm~ observations of the Lockman
Hole. The true $L_{\rm 2-10 keV}$ is generally in  agreement
with the observed $L_{\rm 2-10 keV}$. $M_{\rm BH}$ is derived from
the empirical virial correlation, which mainly depends on the
emission-line width and weakly on the continuum luminosity.}
\end{figure}


\begin{table*}[t1]
\begin{center}
\footnotesize \centerline{\sc Table 1. Low-redshift radio-quiet AGNs
with reverberation-based mass } \vglue 0.1cm
\begin{tabular}{lllllllcc}\hline \hline
Name  & $M_{\rm BH}$ & $\Gamma_{\rm 2-10 keV}$ &$L_{\rm X}$ &$L_{\rm bol}$& $\dot
m_{\rm E}$ & $\kappa_{\rm 2-10 keV}$ & $L_{\rm X}/L_{\rm Edd}$& ref. \\
(1) & (2)    & (3)  & (4)  &(5)&  (6) & (7) & (8) & (9)  \\\hline
Mrk 335   &  14.2$\pm$3.7& $2.29\pm0.02$ &$43.3$&$45.3$&$1.13$&$102^{+4}_{-4}$ & $-1.95$ & 1  \\
PG 0052+251  & $369\pm76$ & $1.81\pm0.04$&$44.6$&$45.8$&$0.148$&$19.5^{+0.7}_{-0.6}$ & $-2.07 $ & 2 \\
F9  &  255$\pm$56  &$1.80\pm0.02$&$43.8$&$44.8$&$0.0186$&$10.5^{+0.8}_{-0.8}$ & $-2.71$ & 3 \\
Mrk 590    &  47.5$\pm$7.4 &$1.66\pm0.03$&$43.0$&$43.8$&$0.0104$&$7.0^{+0.2}_{-0.2}$ &
                 $-2.68$ & 4 \\
Akn 120    &  $150\pm19$ &$2.01\pm0.06$&$43.9$&$45.3$&$0.111$&$25.0^{+0.2}_{-0.3}$ & $-2.38$ & 5 \\
Mrk 79     & $52.4\pm14.4$ &$1.85\pm0.04$&$43.3$&$44.3$&$0.0309$&$10.5^{+1.0}_{-0.9}$&$-2.52$ & 6 \\
PG 0844+349&  $92.4\pm38.1$&$2.11\pm0.05$&$43.6$&$45.4$&$0.233$&$72^{+8}_{-8}$ & $-2.47$ & 7  \\
Mrk 110    &  $25.1\pm6.1$&$1.79\pm0.01$&$43.9$&$45.1$&$0.433$&$18.4^{+0.1}_{-0.1}$ &$-1.60$ & 8 \\
PG 0953+414 & $276\pm 59$&$2.12\pm0.03$&$44.7$&$46.5$&$0.892$&$71^{+2}_{-2}$&$-1.84$ & 9\\
NGC 3227 (1)&  $42.2\pm21.4$&$1.57\pm0.03$&$42.1$&$42.9$&$0.00151$&$7.02^{+0.05}_{-0.05}$&$-3.63$ & 10  \\
NGC 3227 (2)&   $42.2\pm21.4$ &$1.52\pm0.01$&$41.5$&$42.4$&$0.000447$&$8.0^{+0.1}_{-0.1}$&$-4.23$ & 11 \\
NGC 3516  & $42.7\pm14.6$&$1.80\pm0.02$&$42.3$&$43.5$&$0.00612$&$15.32^{+0.10}_{-0.10}$&$-3.43$& 4 \\
NGC 3783 &$29.8\pm5.4$  & $1.73\pm0.01$&$42.9$&$44.1$&$0.0363$&$17.3^{+0.2}_{-0.2}$ & $-2.67$ & 12 \\
NGC 4051 (1)&$1.91\pm 0.78$&$1.90\pm0.02$&$40.8$&$42.6$&$0.0164$&$67^{+4}_{-3}$& $-3.58$ &13 \\
NGC 4051 (2)&$1.91\pm 0.78$&$1.24\pm0.04$&$41.4$&$42.6$&$0.0151$&$15.1^{+0.2}_{-0.1}$&$-2.98$& 13 \\
NGC 4151 & $13.3\pm4.6$&$1.65\pm0.03$&$42.8$&$44.0$&$0.0558$&$15.64^{+0.08}_{-0.08}$& $-2.42$ & 4 \\
PG 1211+143 &  $146 \pm 44$&$1.99\pm0.06$&$43.2$&$45.7$&$0.260$&$340^{+80}_{-60}$&$-3.06$ & 14  \\
PG 1229+204    & $73.2 \pm 35.2$&$1.97\pm0.04$&$43.4$&$44.9$&$0.0823$&$31^{+2}_{-2}$&$-2.56$ & 2 \\
NGC 4593   & $9.8\pm2.1$&$1.83\pm0.02$&$42.8$&$43.7$&$0.0369$&$7.7^{+0.1}_{-0.1}$&$-2.03$ & 15  \\
PG 1307+085 & $440\pm123$&$1.50\pm0.04$&$44.0$&$45.6$&$0.0659$&$35^{+1}_{-1}$&$-2.74$ & 9\\
Mrk 279 & $34.9\pm9.2$&$1.86\pm0.02$&$43.6$&$45.0$&$0.210$&$21.7^{+0.3}_{-0.3}$&$-2.04$ & 4 \\
NGC 5548    & $49.4\pm7.7$     &$1.68\pm0.03$&$43.3$ &$44.3$&$0.0236$&$10.1^{+0.1}_{-0.1}$&$-2.49$ & 4 \\
PG 1426+015 & $886\pm187$&$1.99\pm0.04$&$44.1$&$45.6$&$0.0243$&$30^{+40}_{-10}$&$-2.95$& 16 \\
PG 1613+658 & $279\pm129$&$1.80\pm0.07$&$44.1$&$45.9$&$0.221$&$67^{+20}_{-10}$&$-2.45$ & 9\\
Mrk 509     & $143\pm12$  &$1.72\pm0.02$&$44.0$&$45.2$ &$0.0951$&$16.20^{+0.10}_{-0.09}$&$-2.26$ &17 \\
PG 2130+099 & $38\pm15$&$1.65\pm0.03$&$43.5$&$45.0$&$0.0179$&$35^{+1}_{-1}$&$-2.18$ & 18\\
NGC 7469    & $12.2\pm 1.4$&$1.75\pm0.01$&$43.2$&$44.8$&$0.369$&$42^{+1}_{-1}$&$-1.99$& 19 \\\hline
\end{tabular}
{\baselineskip 8pt \indent\\
(1) Object name. (2) Black hole mass, in units of $10^{6}M_{\odot}$,
taken from Peterson et al. (2004), except for NGC 4593 (Denney et
al. 2006) and PG 2130+099 (Grier et al. 2008). (3) Rest-frame 2-10
keV photon index. (4) Rest-frame 2-10 keV luminosity. (5) Bolometric
luminosity. (6) Eddington ratio, $\dot m_{\rm E} \equiv L_{\rm
bol}/L_{\rm Edd}$. (7) Bolometric correction, $\kappa_{\rm 2-10
keV}\equiv L_{\rm bol}/L_{\rm 2-10 keV}$. (8) Ratio of rest-frame
$2-10$ keV luminosity to the Eddington luminosity. (9) Reference for
the X-ray observation. (1) Gondoin et al. (2002); (2) D'Ammando et
al. (2008); (3) Gondoin et al. (2001); (4) Zhou \& Zhang (2010); (5)
Vaughan et al. (2004); (6) Gallo et al. (2005); (7) Brinkmann et al.
(2003); (8) Dasgupta \& Rao (2006); (9) Piconcelli et al. (2005);
(10) Markowitz et al. (2009); (11) Gondoin et al. (2003); (12)
Reeves et al. (2004); (13) Ponti et al. (2006); (14) Reeves et al.
(2005); (15) Steenbrugge et al. (2003); (16) Page et al. (2004);
(17) Ponti et al. (2009); (18) Inoue et al. (2007); (19) Blustin et
al. (2003).}
\end{center}
\end{table*}

\end{document}